\documentclass{ws-mpla}

\newcommand{\babar}{\mbox{\sc{BaBar}}}
\newcommand{\coupling}{\kappa}

\begin{document}

\markboth{C. Bird, R. Kowalewski, M. Pospelov}
{Dark matter pair-production in $b \to s$ transitions}

\catchline{}{}{}{}{}

\title{Dark matter pair-production in $b \to s$ transitions.}

\author{C. Bird\footnote{cbird@uvic.ca},\: R. Kowalewski\footnote{kowalews@uvic.ca},\; 
M. Pospelov\footnote{pospelov@uvic.ca}}

\address{Department of Physics and Astronomy, University of Victoria, Victoria, BC, V8P 1A1, Canada
}


\maketitle

\pub{Received (Day Month Year)}{Revised (Day Month Year)}

\begin{abstract}

The flavour-changing neutral current transition $b \to s$ can serve as 
a sensitive probe of WIMP dark matter models, if the WIMP mass
is under 2 GeV. In this work we extend our earlier analysis
to a generic class of models where the interaction between 
 the dark matter sector and the Standard Model matter sector is mediated 
 by the Higgs boson(s). We show that experimental limits on the decays of 
  $B$-mesons to $K$($K^*$) and missing energy provide stringent constraints on 
  the parameter space of such models, but do not 
  rule 
out sub-GeV WIMPs in a model-independent way. We find that in the context of the 
  NMSSM with light pseudoscalar Higgs, the WIMP masses under a few hundred MeV 
  are generically excluded with the exception of few highly tuned 
  points in the parameter space. 

\end{abstract}

\ccode{PACS Nos.: include PACS Nos.}

\section{Introduction}	

\par
The dark matter content of the total energy density in the 
Universe is now well-measured, with several experiments determining an abundance
 of $\Omega_{DM} h^2 \sim 0.12$ \cite{Spergel:2003cb}. So far the presence of 
dark matter was deduced only through its gravitational interaction, and thus its identity remains a
mystery.

\par
Among numerous candidates for dark matter suggested in the literarture, the weakly interacting 
massive particles (WIMPs) are the most interesting from the point of view of particle
 physics. 
The residual abundance of such particles in the early Universe is 
well understood, with the observed dark matter abundance providing a
  measure of the WIMP couplings through their annihilation cross section 
at freeze-out (for a review, see Ref. \refcite{Kolb:1990vq}). Furthermore the inferred
   couplings of WIMPs to the Standard Model fields are sufficiently large that detection 
of WIMP pair-production via a missing energy
   signal is possible in current particle physics experiments.
   
\par
The mass of the WIMP can be constrained from above and below using certain theoretical arguments.
The assumptions that the mass scale of the particles that mediate interactions between 
Standard Model and WIMPs is not lighter than the weak scale, and couplings are perturbative
produce the WIMP annihilation cross sections that vanish both in the limits of very large and 
very small WIMP masses. Comparison with 
$\sigma_{ann} v_{rel} \sim 0.1 \; pb$ , which is required to reduce the dark matter abundance to an acceptable level during
 freeze-out, produces the allowed mass range, referred to as the Lee-Weinberg window \cite{Lee:1977ua,Russians}.
Obviously, such a window is very model dependent.  In many popular models this limit 
requires the WIMP mass to be larger than a few GeV. However it is possible to construct 
simple models in which the required masses can be as low as 100 MeV
   (e.g. Ref \refcite{Burgess:2000yq,Fayet,Fayet2}). The Lee-Weinberg limit is usually 
quoted for fermions, in which case the annihilation cross-section is suppressed by
    a factor of $m_{DM}^2/ M^2$, where $M$ is the mass of the mediator particle, 
which requires large couplings for light WIMPs. However if the cross section is
     enhanced, or if scalars are used instead of fermions, the lower limit on
WIMP masses can be significantly reduced.

\par
\begin{figure}
\begin{center}
\psfig{file=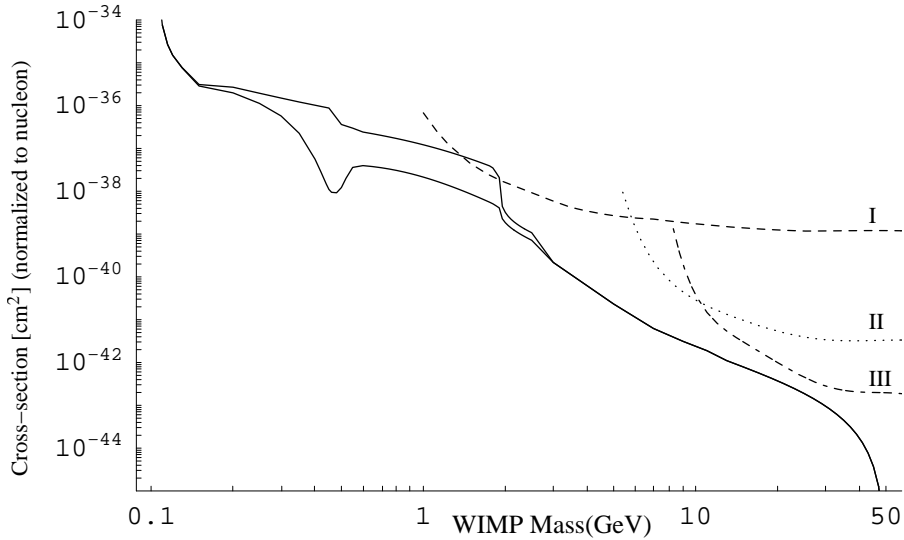,width=12cm,angle=0}
\end{center}
\caption{\label{figure:Limits} Current limits on WIMP-nucleon spin-independent 
cross sections from dedicated underground searches. The solid lines represent 
the predictions for the minimal scalar model with a $100  $ GeV Higgs, 
while the current limits are given from (I) CRESST, (II) CDMS (2005 Si), and (III) CDMS (2005 Ge). 
In the interval of 100 MeV - 2 GeV the predicted signal has signiciant QCD-related uncertainty.}
\end{figure}

Light dark matter models have several interesting features. The dedicated dark matter searches, 
such as DAMA and CDMS, rely on measuring the recoil of nuclei from WIMP scattering and as 
such are less sensitive to WIMPs which are significantly lighter than the nuclei used. Light WIMPs produce 
small recoil energy, which is usually well under the background that peaks at small $E$.  
As indicated in Figure \ref{figure:Limits} this insensitivity to sub-GeV WIMPs 
will not be solved with the next generation of underground experiments, 
although the CRESST experiment is expected to provide an improved probe of dark matter as light as $O(1 $ GeV)
due to its low recoil threshold.
Interestingly enough, there is a possibility
that a light WIMP could explain the positive signal observed in the 
annual modulation at DAMA without violating the bounds from other experiments \cite{Gondolo:2005hh}. 

\par
It has also been suggested that the annihilation of light WIMPs could explain the 
galactic positron excess \cite{Boehm:2003bt}. 
The positron excess has been measured through the flux of 
$511 \; {\rm KeV} \; \gamma$-rays meassured by the INTEGRAL/SPI experiment 
\cite{Knodlseder:2003sv,Jean:2003ci}, and seems to require non-standard sources of positrons. Although there are
 other possible astrophysical explanations 
(see for example Ref. \refcite{altex,altex2,altex3}), 
positron production by annihilations \cite{Boehm:2003bt} or decays of light dark matter 
particles \cite{Picciotto:2004rp} remains as one of the more 
attractive solutions due to its natural explanation of the nonlocalized nature of 
the observed $\gamma$-rays.  It has also been demonstrated that the 
annihilation of light dark matter does not conflict with measurements of either the galactic $\gamma$-ray flux \cite{Beacom:2004pe,Beacom:2005qv} or the extragalactic
   $\gamma$-ray background\cite{Rasera:2005sa} , and may also provide an 
explanation for the observed flux of $\gamma$-rays in the 1-20 MeV range\cite{Ahn:2005ck}.

\par
As we recently demonstrated, an intriguing possibility of a sub-GeV WIMPs is
 the opportunity of detecting a pair of WIMPs as a missing energy signal in 
$B$ meson decays \cite{Bird:2004ts}. 
Indeed, as dedicated underground searches for dark matter remain largely insensitive to 
sub-GeV WIMPs, 
their production is possible in experiments such as \babar\ and BELLE, which
 produce large numbers of $B$-mesons, and can study their rare
decay modes. 
As a result such facilities provide a new opportunity to search for light dark matter. 
For the minimal scalar WIMP model these experiments have already excluded most of the
 parameter space with $m_S \lesssim 1 $ GeV, while future data from $B$ factories 
will be able to probe as high as $m_S \sim 2$  GeV \cite{Bird:2004ts}. 

The purpose of this paper is to explore the question of how  generic the limits on light WIMPs 
derived in Ref. \refcite{Bird:2004ts} are, and whether all dark matter models with sub-GeV 
WIMPs can be efficiently constrained by $B$-physics.
To answer these questions we study the class of models where the interaction between 
Standard Model sector and WIMPs is mediated by one or more Higgs particles. 
 We demonstrate that $b\to s$ decays with missing energy 
provide  important constraints on the parameter space of such models. We also 
point out the 
possibility, based on the two-Higgs doublet model (2HDM) at 
large $\tan\beta$, that these constraints can be circumvented. 

In Section
 \ref{MSM} we review our previous results on the minimal scalar model and 
extend the result for more general scalar models with an additional singlet scalar that 
mixes with the Higgs boson. In Section \ref{2HDM} we apply the 
same tecniques to a related model with two Higgs doublets and calculate the 
branching ratios of WIMP-producing decays of $B$-mesons. This model has the 
additional benefit of relaxing the fine tuning 
condition required for a sub-GeV scalar WIMP
in the minimal model. In Section \ref{Fermion} we introduce some 
simple models of fermionic dark matter, calculate the WIMP production in $B$-decays, and 
discuss the limitations on such models from the Lee-Weinberg limit. 
We also address the case of NMMSM (next-to-minimal supersymmetric Standard Model)
where we show that WIMPs under 200 MeV are generically excluded 
except special points in the parameter space with $b\to s$ transitions being fine-tuned
well below its natural scale.
In Section \ref{experiment} we review the
    experimental bounds on the decay $B \to K + missing \; energy$, and discuss the potential for future searches. 
    
\par
Before we present our calculations, we would like to make 
several important comments.
In our discussions of the Lee-Weinberg limit, it should be noted that 
there is no clear division between the perturbative and non-perturbative 
regions of parameter space and therefore no definitive lower bound on the WIMP mass. 
However in this paper we will require the WIMPs to be sufficiently heavy 
that the abundance constraints satisfy $\kappa^2 \lesssim 4 \pi$, with $
\kappa$ representing an effective coupling in each model. 

\par
As it is done in Ref. \refcite{Bird:2004ts}, we treat the significant hadronic uncertainties 
in the annihilation cross section using the most "optimistic" and "pessimistic" scenarios. That
results in the certain range for predicted WIMP-producing branching ratios.
In most of the models we present, the largest uncertainty exists
 for a WIMP with mass $m_{DM} \sim 500 $ MeV that
could annihilate through the $f_0$ resonance potentially enhancing the cross section 
and resulting in weaker couplings.  However, it is unclear 
   whether such resonances exist at the freeze-out temperature of $T \sim O(50$ MeV). 
The heavier resonances are not expected to contribute to the
    annihilation cross section as the higher freeze-out 
temperatures will significantly reduce the strength of the resonances. 
We also will assume that WIMPs heaver than $\sim 1 \;$GeV 
will freeze-out before the transition from quarks to hadrons.

\par
We will also limit our discussion to the decay channel 
$B \to K + missing \; energy$. While it is possible to produce WIMPs in 
other decays, such as $B \to missing \; energy$ or $B \to \gamma + missing \; energy$,
the experimental limits on such decays are inferior to $B \to K + missing \; energy$. 
Heavier WIMPs could be produced in $\Upsilon \to \gamma + missing \; energy$,
however a SM decay of $\Upsilon$ is due to strong interactions, which 
greatly reduces the branching ratio of WIMP production compared to weak $B$-meson decays.

\par
Finally, it should  be noted that the collection of models 
presented in this paper is not exhaustive. There are  other possibilities
which could result in light dark matter \cite{Fayet,Fayet2,NMSSM}, 
that generically require relatively light particle mediating the 
interaction between the dark matter sector and the SM sector.  
In this paper, we will breifly touch a possibility of a 
lighter pseudoscalar Higgs particle discussed in Ref. \refcite{NMSSM},
and omit the models with an extra $U(1)$ gauge group 
that have an additional $Z'$ in the sub-GeV range. 
The latter possibility is interesting, and 
should such $Z'$ have axial-vector couplings to fermions, 
one expects large flavour-changing $b-s-Z'$ amplitudes 
induced by radiative corrections. However, should this 
new force couple to a pure vector current (with {\em e.g.} $B-L$
quantum numbers), the constriants from flavour-changing decays will 
be greatly relaxed. The detailed analysis of models with light WIMPs supplemented by light $Z'$ 
goes outside the scope of the present paper.

\section{\label{MSM} Minimal Scalar Models }

\par
The simplest WIMP model is a singlet scalar\cite{Silveira:1985rk,McDonald:1993ex,Burgess:2000yq} 
which interacts with the Standard Model through exchange of the Higgs:

\begin{equation}
\label{lagr}
\begin{split}
-{\cal L}_S &= \frac{\lambda_S}{4}S^4+ \frac{m_0^2}{2} S^2+ 
\lambda S^2  H^\dagger H \\&\\&= \frac{\lambda_S}{4}S^4+
\frac12 (m_0^2 + \lambda v_{EW}^2) S^2 + \lambda v_{EW} S^2  h + \frac{\lambda}{2}S^2 h^2,
\end{split}
\end{equation}

\noindent
where $H$ is the SM Higgs field doublet, $v_{EW} = 246$ GeV is the 
Higgs vacuum expectation value (vev) and $h$ is the corresponding physical 
Higgs, $H = (0,(v_{EW}+h)/\sqrt 2)$. The physical 
mass of the scalar $S$ receives contributions from two terms,  
$m_S^2 = m_0^2 + \lambda v_{EW}^2$,
and requires significant fine-tuning to provide a sub-GeV mass. In this section
we will calculate the branching ratio for the pair production of scalars
in the decay $B \to K + SS$, which contributes to Br$(B^+ \to K^+ + missing \; energy)$.
Being minimal, this model obviously possesses maximum predictivity, and 
the branching ratio of WIMP production can be calculated as a function 
of dark matter mass only. 

\begin{figure}
\psfig{file=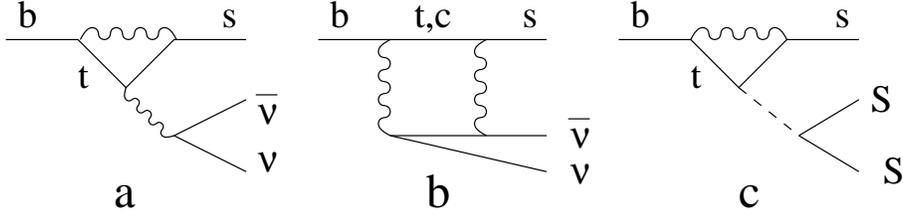,width=12cm,angle=0}
\caption{\label{diagram:MSMgraphs}Feynman diagrams which contribute to B-decay with missing energy in the minimal scalar model of dark matter.}
\end{figure}

\par
It should be noted that the decay $B \to K + missing \; energy$ is actually expected to occur regardless 
of the existence or nature of light dark matter. As shown in Figure \ref{diagram:MSMgraphs}a and 
\ref{diagram:MSMgraphs}b, the Standard Model predicts the transition $b \to s + \overline{\nu} \nu$ at one loop, 
so that the $B$-meson can decay to neutrinos  \cite{Buchalla:2000sk}, with 
Br$(B^+ \to K^+ + missing \; energy) \simeq (4\pm1) \times 10^{-6}$.
 However as
  demonstrated before \cite{Bird:2004ts}, the decay $B \to K + SS$ (resulting from the $b\to s$ transition shown in Figure
   \ref{diagram:MSMgraphs}c) can enhance the missing energy signal by up to two orders of magnitude.

\par
The transition $b \to s +h$ occurs as a loop process, which at low momentum 
transfer can be calculated by differentiation of the $b \to s$  self-energy operator with
respect to $v_{EW}$,

\begin{equation} 
{\cal L}_{bsh} = 
\left(\frac{3 g_W^2 m_b m_t^2 V_{ts}^* V_{tb}}{64 \pi^2 M_W^2 v_{EW}} 
\right) \overline{s}_L b_R h +(h.c.).
\label{bsh}
\end{equation}

\noindent
As the Higgs is significantly heavier than the other particles 
involved in the process, it can be integrated out leaving an 
effective Lagrangian for the $b \to s$ transitions with missing energy:

\begin{equation}
{\cal L}_{b\rightarrow s E\!\!\!\!/} = \frac{1}{2}C_{DM} m_b\bar s_L b_R S^2 -
C_\nu \bar s_L \gamma_\mu b_L \bar \nu \gamma_\mu \nu+(h.c.). 
\label{bsE}
\end{equation}

\noindent
Leading order Wilson coefficients for the transitions with dark matter scalars or 
neutrinos in the final state are given by 
\begin{eqnarray}
\label{CDM}
C_{DM} = \frac{\lambda }{m_h^2} ~\frac{ 3  g_W^2 V_{ts}^* V_{tb}}{32 \pi^2}~x_t
\;\;\;\;\;\;\;\;\;\;\;\;\\
C_\nu = \frac{g_W^2 }{M_W^2} ~\frac{g_W^2V_{ts}^* V_{tb}}{16\pi^2}
\left[\frac{x_t^2+ 2 x_t}{8(x_t-1)}
+\frac{3x_t^2-6x_t}{8(x_t-1)^2}\ln x_t \right],
\nonumber
\end{eqnarray}
\noindent
where $x_t = m_t^2/M_W^2$. From this effective theory and the hadronic 
form factors calculated in the light-cone sum rules \cite{Ali:1999mm,Bobeth:2001sq} , 
the partial width for decays with missing energy can be calculated

\begin{eqnarray}
{\rm Br}_{B^+\to K^+ + E\!\!\!\!/}= {\rm Br}_{B^+\to K^+\nu\bar\nu} + 
{\rm Br}_{B^+\to K^+SS} \nonumber 
\\
\simeq 4\times 10^{-6} + 2.8 \times 10^{-4} \kappa^2 F(m_S).
\label{totalbr}
\end{eqnarray}

\noindent
where we use the parametrization 
\begin{equation}
 \coupling^2 \equiv \lambda^2 \left( \frac{100 ~{\rm GeV}}{m_h} \right)^4,
\end{equation}

\noindent
and the available phase space as a function of the unknown $m_S$, 
\begin{displaymath}
F(m_S)=\int_{\hat s_{min}}^{\hat s_{max}}\!\!\!\!\!\!\! 
f_0(\hat{s})^2 I(\hat s,m_S)~d\hat s ~\left[
\int_{\hat s_{min}}^{\hat s_{max}} \!\!\!\!\!\!\! 
f_0(\hat{s})^2 I(\hat s,0)~d\hat s~\right ]^{-1},
\end{displaymath}
with
$$
I(\hat s,m_S)= [\hat s^2 - 2\hat s(M_B^2 +M_K^2)+(M_B^2-M_K^2)^2]^{\frac{1}{2}}
[1-4m_S^2/\hat s]^{\frac{1}{2}}.
$$

\begin{figure}
\label{figure:BranchRatio}
\begin{center}
$\begin{array}{c}
\psfig{file=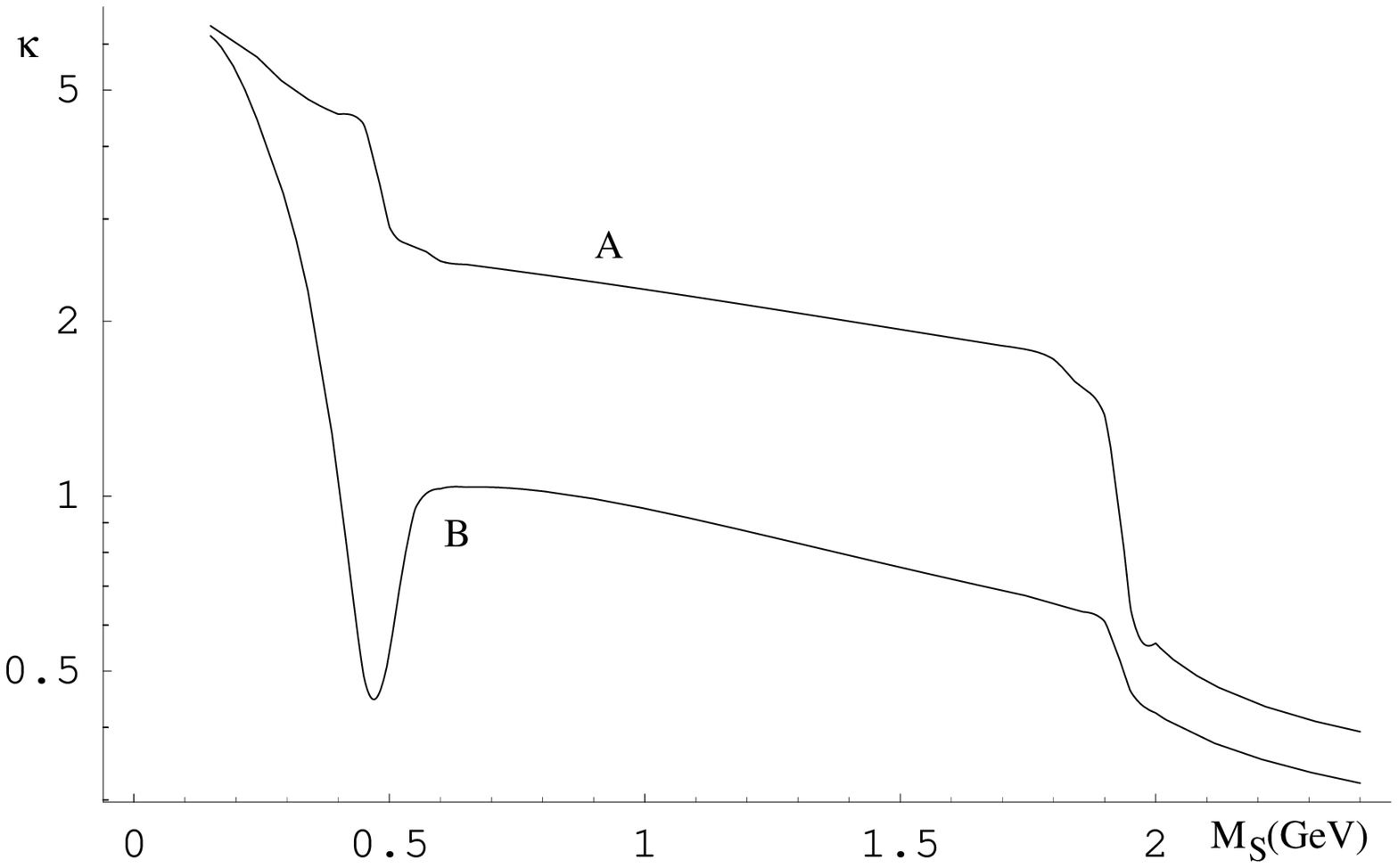,width=11cm,angle=0} \\(a) \\
\psfig{file=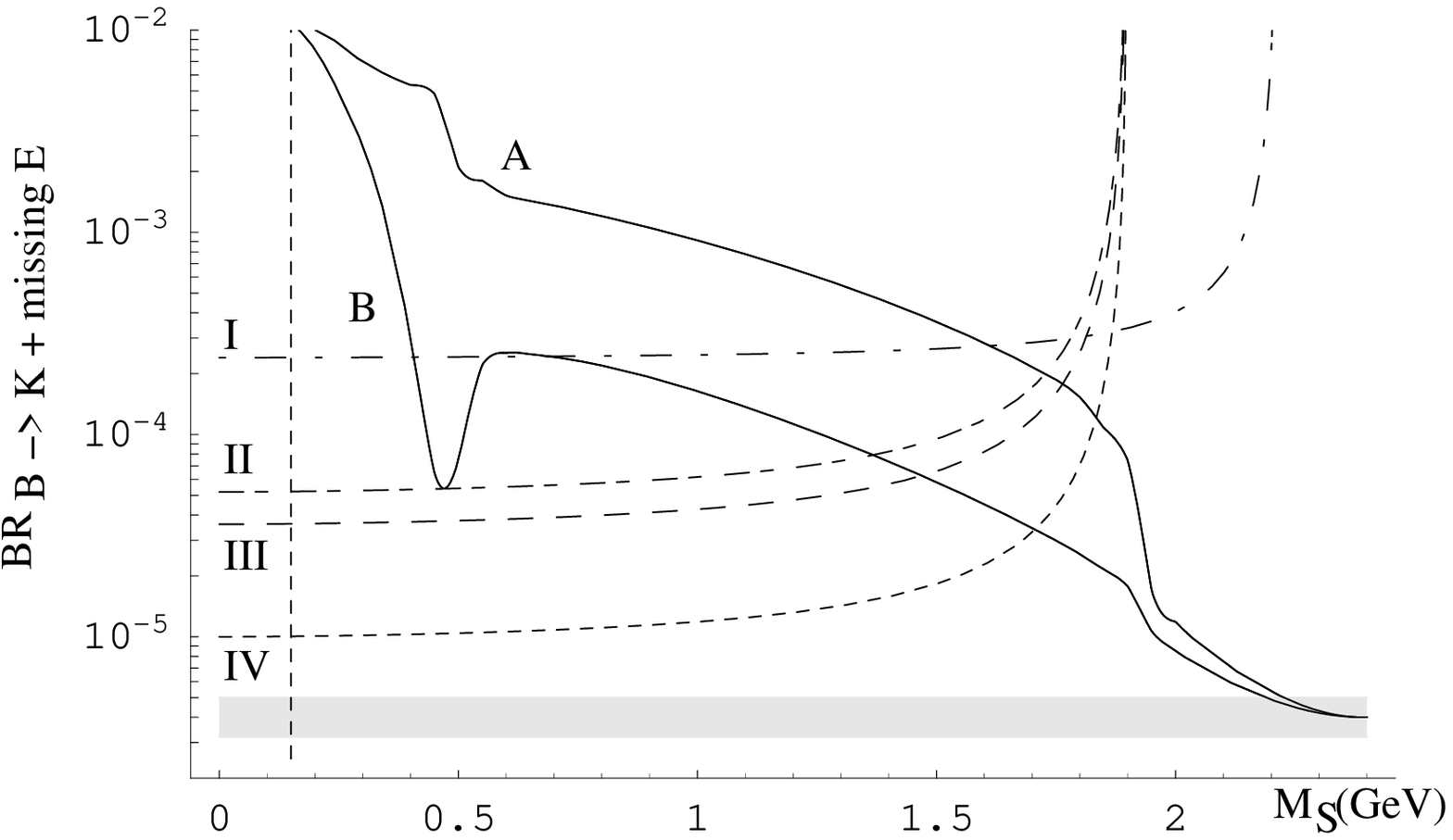,width=11cm,angle=0} \\ (b)
\end{array}$
\end{center}
(a) Constraints on $\kappa$ from dark matter abundance. 
The parameter space above A produces a density too small to explain dark matter, 
while the space below B would lead to a higher than observed dark matter abundance.
(b) Predicted branching ratios for 
the decay $B^+ \to K^+ +$ {\it missing energy}, with current limits 
from CLEO (I) \cite{Browder:2000qr} , $\babar$  (II) \cite{Aubert:2004}, Belle (III) \cite{Abe:2005bq}
and expected results from $\babar$ (IV). Parameter space above curves I, II and III is excluded.
The grey bar shows the expected $B\to K\nu\bar \nu$ signal. Parameter 
space to the left of the vertical dashed line can also be probed with  $K^+ \to \pi^+ + missing ~energy $.
\end{figure}

\par For comparison, \babar\ recently reported\cite{Aubert:2004} a
limit of Br$(B^+ \to K^+ +missing \; energy) < 5.2 \times 10^{-5} $ at
$90\%$ c.l. which is well above the SM prediction.  Belle's
preliminary results were reported at Br$(B^+ \to K^+ +missing \;
energy) < 3.5 \times 10^{-5} $ level \cite{Abe:2005bq}.  Similar
calculations can be used for the decay $B \to K^* SS$,

\begin{equation}
{\rm Br}_{B^+\to K^{+*} + E\!\!\!\!/}
\simeq 1.3\times 10^{-5} + 3.0 \times 10^{-4} \kappa^2 F(m_S).
\end{equation}

\noindent
with an analogous form factor.

Even before we perform   the freeze-out abundance calculation and
extract $\kappa$, it is obvious that the decay into WIMPs would dominate the
Standard Model rate to neutrinos by one-to-two orders of magnitude near 
the Lee-Weinberg bound of $\kappa \sim O(1)$. 
To make this statement more precise, we extract the parameter $\kappa$ using the relation 
of the WIMP annihilation cross section, 

\begin{equation}
\sigma_{ann} v_{rel} = \frac{8 v_{ew}^2 \lambda^2}{m_h^4} \times \lim_{m_{\tilde{h}} \to 2 m_S } 
\left( \frac{\Gamma_{\tilde{h} \to X}}{m_{\tilde{h}}} \right) 
\end{equation}

\noindent
with the observed dark matter abundance \cite{Spergel:2003cb},
 $\Omega_{DM}h^2 \sim 0.12$. Here $\Gamma_{h \to X}$  is the total decay width for a
 virtual Higgs of mass $m_{\tilde{h}} \sim 2 m_S$, and in our calculation we 
utilize the zero temperature decay widths which were studied previously in 
early searches for a light Higgs \cite{Voloshin:1985tc,Raby:1988qf,Truong:1989my} 
as explained in our previous paper \cite{Bird:2004ts}. 
The allowed range of $\kappa$ is plotted in Figure
  \ref{figure:BranchRatio}a, where the main uncertainty 
  comes from $\Gamma_{h \to hadrons}$. It should be kept in mind that the domain of $\kappa$'s
  {\em below} curve B gives an over-production of dark matter and therefore is
  firmly excluded. Inserting $\kappa-m_S$ domain allowed by the relic abundance
into the result for the branching ratio gives the prediction of Br$(B^+ \to K^+ +missing \; energy)$
as a function of $m_S$, Figure \ref{figure:BranchRatio}b. 
  
\par
These results can be easily generalized to models of scalar dark matter
coupled to the SM via an additional singlet Higgs particle $U$.
The simplest model of this type has the potential

\begin{equation}
\begin{split}
-{\cal L}_S &= \frac{\lambda_S}{4}S^4+\frac{m_0^2}{2} S^2+
(\mu_1 U + \mu_2 U^2) S^2   + V(U)+ \eta' U^2 H^{\dag} H 
\\&\\
&= \frac{m_S^2}{2} S^2+\frac{m_u^2}{2} u^2+ 
\mu u S^2   + \eta v_{EW} u h +...,
\end{split}
\label{tsdm}
\end{equation}
where in the second line we retained only mass terms and relevant interaction terms.
$u$ denotes the excitation around the vev of $U$ and $\mu$ and $\eta$ stand for 
dimensionful parameters presumably of order the electroweak scale. 
\noindent
 The last term in the second line of (\ref{tsdm})
gives the mixing between scalars $u$ and $h$. If such mixing is significant,
the existing bounds on the higgs mass would also place a lower bound on 
the mass of the $u$-boson. 
The effective Lagrangian for $b \to s+E\!\!\!\!/$ transitions is the 
same as Eq. \ref{bsE}, with the Wilson coefficient given by

\begin{equation}
C_{DM} = \frac{\mu \eta  }{m_u^2 m_h^2} \frac{ 3  g_W^2 V_{ts}^* V_{tb}}{32 \pi^2}x_t
\end{equation}

\noindent
which, with the redefinition 

\begin{displaymath}
\kappa^2 \equiv \frac{\mu^2 \eta^2 }{m_u^4} \left(\frac{100 \;{\rm GeV}}{m_h} \right)^4
\end{displaymath}

\noindent
results in the same abundance constraints and the same 
branching ratio for $B\to K + missing \; energy$, 
as plotted in Figure \ref{figure:BranchRatio}. Thus, we see that the 
model with more parameters in the singlet sector (\ref{tsdm}) in the limit of light WIMPs gives
identical predictions to the minimal model (\ref{lagr}), and in this sense 
the predictions of Figure \ref{figure:BranchRatio}b are generic. 

\section{\label{2HDM} Scalar dark matter in 2HDM Model }

\par
The singlet scalar model of WIMPs coupled to the SM is the most economical model of dark matter, 
and for sub-GeV WIMPs can be well constrained with existing $B$-factory experiments. 
However, the light masses of WIMPs in this model are not natural, as it requires 
significant fine-tuned cancelation between $m_0^2$ and $\lambda v^2_{EW}$. 
However, if the electroweak sector of the SM is modified, this fine-tuning can be
significantly relaxed.

\par
In this section, we consider a singlet scalar WIMP which interacts 
with two Higgs doublets, $H_u$ and $H_d$.

\begin{equation}
-{\cal L} = \frac{m_0^2}{2}S^2 + \lambda_1 S^2 (|H_d^0|^2 + |H_d^-|^2) + 
\lambda_2 S^2 (|H_u^0|^2 + |H_u^+|^2) + \lambda_3 S^2 (H_d^-H_u^+ - H_d^0 H_u^0)
\end{equation}

\noindent
We assume the most conservative flavour arrangement
in which the up- and down-types of quarks originated from the expectation values of different Higgses, 
exactly as it happens in supersymmetric models (see {\em e.g.} Ref. \refcite{Martin:1997ns}). 
This model differs from the SM in a significant way: due to two different vev's of the two Higgs doublets, 
there is an additional parameter $\tan \beta \equiv v_u/v_d$, which is large when Yukawa 
couplings in the down-sector are enhanced. Indeed, there are many theories which 
attempt to unify the Yukawa couplings of the third generation of
 the Standard Model (eg see Ref. \refcite{LargeTanBeta,LargeTanBeta2,LargeTanBeta3}). 
Such unification typically requires 
$\tan \beta \approx m_t/m_b$ or $\tan \beta \approx m_t/m_{\tau}$ and 
thus $v_d\sim O$(few GeV). Therefore in the regime of large $\tan\beta$ the mass corrections which depend only on 
$v_d$ will in general be small and may lead to minimal fine tuning.
\par
For example, the physical mass of the WIMP scalar in the model given above is

\begin{displaymath}
m_S^2 = m_0^2 +  \lambda_1 v_d^2 +  \lambda_2 v_u^2 - \lambda_3 v_u v_d.
\end{displaymath}

\noindent
If $\tan\beta \sim O(1)$, sub-GeV WIMPs would require the same amount of 
fine tuning as before. However in the special case of large $\tan\beta$ 
and the hierarchy of $\lambda$'s,
$\lambda_1 \gg \lambda_2,\lambda_3,$ the mass correction depends only on $v_d$, 
resulting in $\frac{\delta m_S^2}{m_S^2} \sim O(1-10)$ for a GeV-scale  WIMP mass,
which then does not require significant fine tuning. 
In the case of $\lambda_3 \gg \lambda_1,\lambda_2$, the fine-tuning is also relaxed,
especially if the abundance constraint 
allows for $\lambda_3 \sim O(\tan^{-1} \beta)$. 
In the rest of this section, we analyze the 
WIMP production in the model, assuming large $\tan\beta$.

\subsection{$\lambda_1$ dominant}

\begin{figure}
\psfig{file=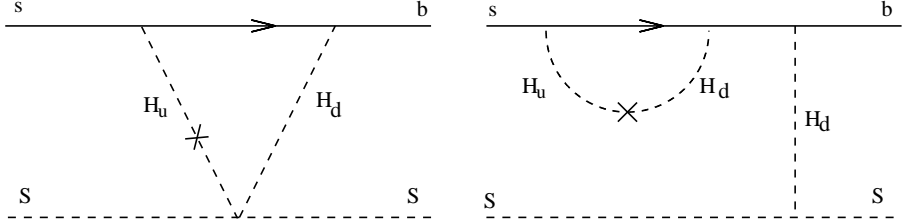,width=12cm,angle=0}
\caption{\label{figure:2HDMdiagrams}Diagrams contributing to the decay $b \to s +SS$ in   the 2HDM plus 
scalar dark matter model when $\lambda_1$ is dominant and $\tan\beta$ is large. 
Inside the loops, $H_u$ and $H_d$ denote the two charged Higgs bosons, 
with the mixing of the two doublets denoted by a cross.}
\end{figure}

\par

The additional diagrams which contribute to the decay $B \to K + SS$ and give the 
leading-order contributions in $\tan\beta$ are 
given in Figure \ref{figure:2HDMdiagrams}. The effective Lagrangian 
is of the same form as Eq \ref{bsE}, and in the limit of 
large $\tan \beta$ and $M_{H_d}^2 \gg M_W^2$ the Wilson coefficient is

\begin{equation}
C_{DM} = \frac{\lambda_1}{M_{H_d}^2} \frac{g_W^2 V_{ts}^* V_{tb} x_t}{32 \pi^2  } 
\left( \frac{1-a_t + a_t \ln a_t}{(1-a_t)^2} \right)
\end{equation}

\noindent
where $a_t = m_t^2/M_H^2$. The branching ratio is calculated as before,

\begin{equation}
{\rm Br}_{B \to K + E} = 4.0 \times 10^{-6} + 3.2 \times 10^{-5} \coupling^2 \left( \frac{1-a_t + a_t 
\ln a_t}{(1-a_t)^2} \right)^2 F(m_S)
\end{equation}

\begin{figure}
\begin{center}
$\begin{array}{c} 
\psfig{file=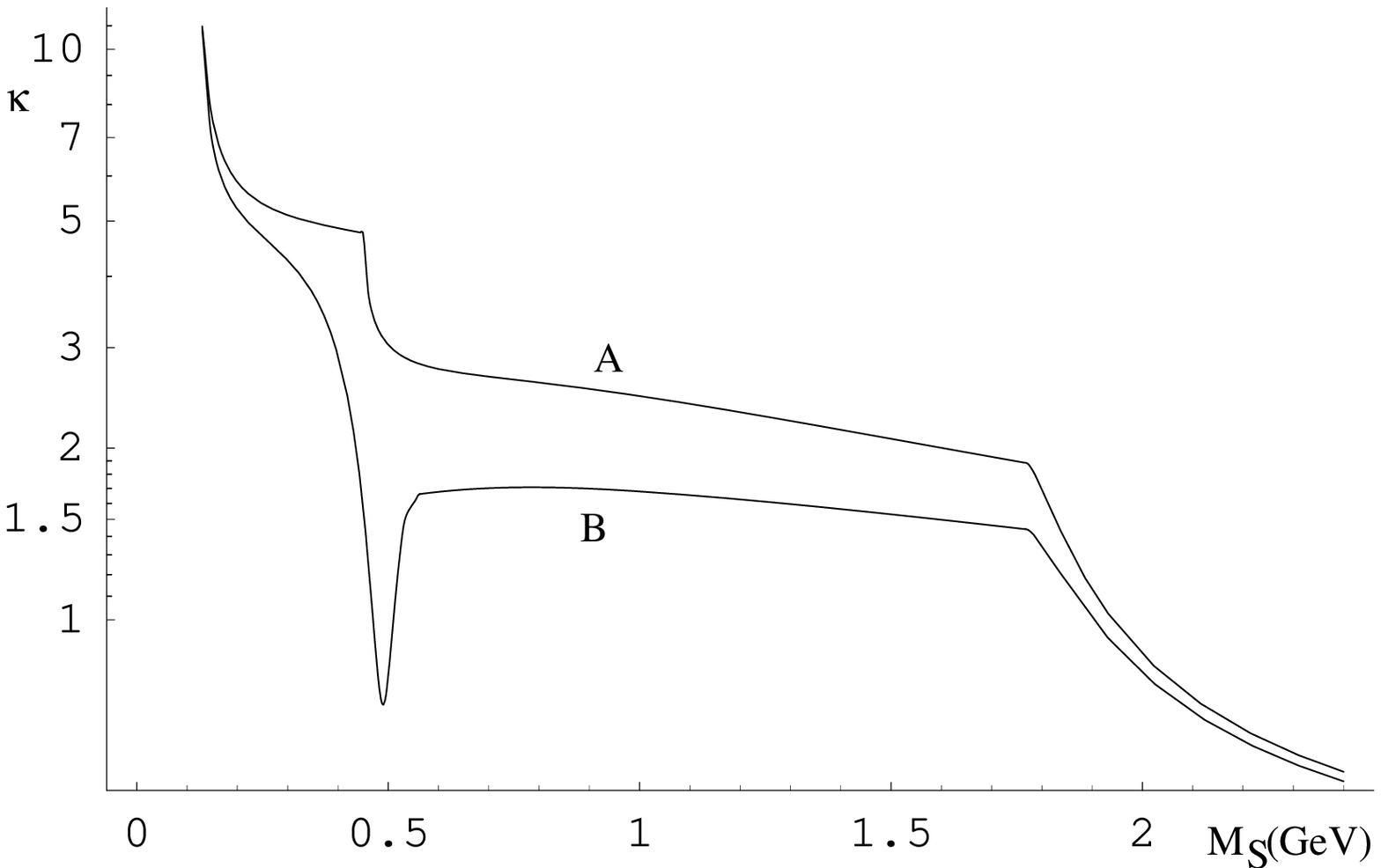,width=12cm,angle=0} \\(a)\\
\psfig{file=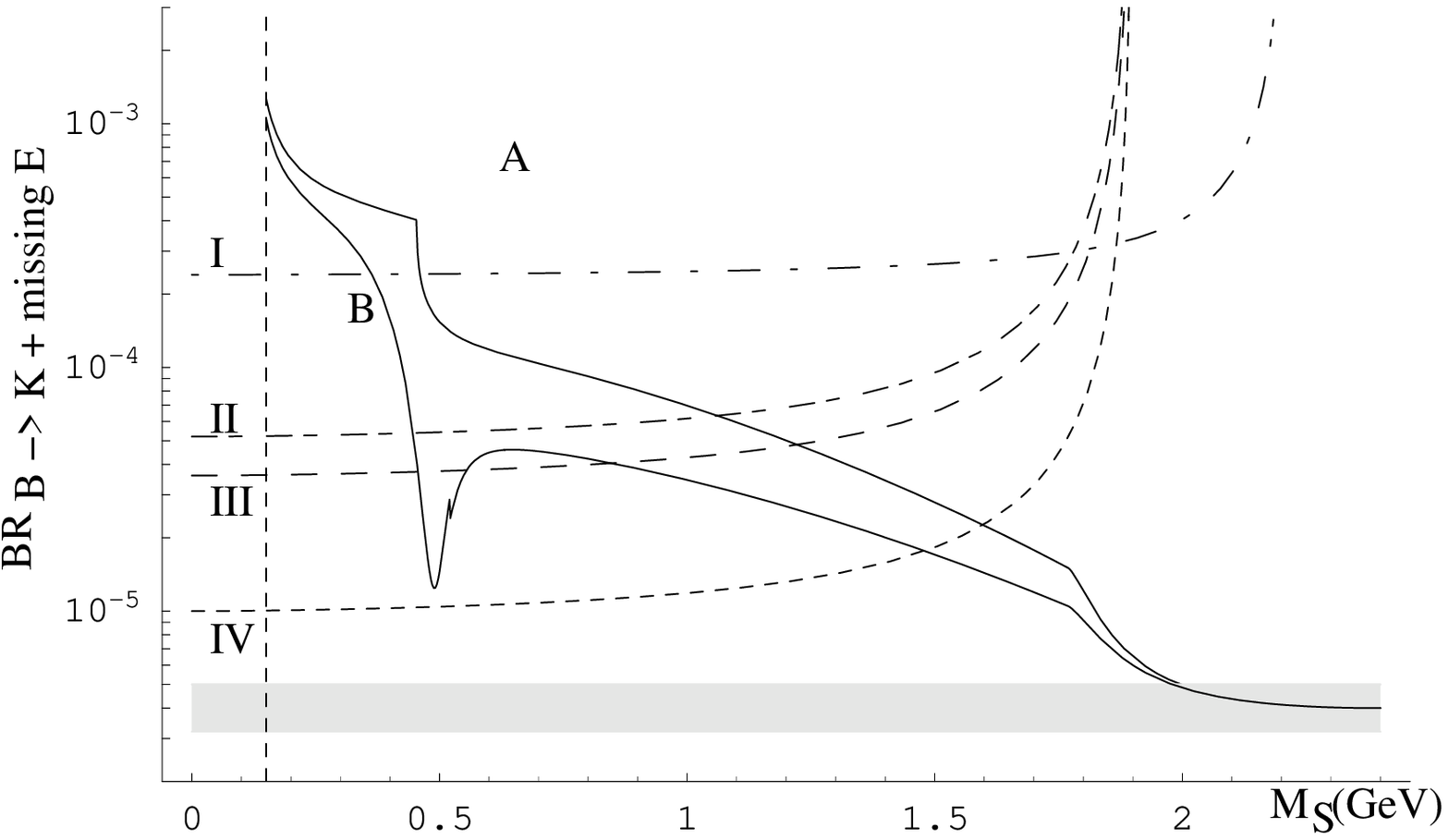,width=12cm,angle=0} \\
 (b)
\end{array}$
\end{center}
\caption{\label{2HDM:BR1}(a) Constraints on $\kappa$ from the observed dark 
matter abundance (b) Branching Ratios for $B \to K+missing \; energy$ in the two higgs doublet model, with WIMPs coupled
 primarily to $H_d$.  The labeling of current limits from \babar,\ BELLE,
and CLEO is the same as
 as in Figure \ref{figure:BranchRatio}b. }
\end{figure}

\noindent
Here we use the parameterization,

\begin{displaymath}
\coupling = \lambda_1 \left( \frac{100 \; GeV}{M_H} \right)^2
\end{displaymath}

\noindent
and unlike the minimal case there is an additional rather mild dependence on $m_H$ through $a_t$. 
It disappears in the limit $m_H\gg m_t$. We also make a safe assumption that additional diagrams 
with charged Higgses are not going to alter the branching to neutrinos.  
As before, $F(m_S)$ is constructed such that
\begin{displaymath}
F(m_S) = \int_{s_{min}}^{s_{max}} f_0(s)^2 I(s,m_S) ds \left[ \int_{s_{min}}^{s_{max}} f_0(s)^2 I(s,0) ds \right]^{-1}
\end{displaymath}
and
 $F(0)=1$, and $F(m_S) = 0$ for $m_S > \frac{1}{2} (m_B-m_K)$. 
The branching ratio is plotted in Figure \ref{2HDM:BR1}b with the current experimental bounds 
from \babar\ \cite{Aubert:2004}, BELLE \cite{Abe:2005bq} and  CLEO \cite{Browder:2000qr}. 

\par
The abundance constraint on $\kappa$ is calculated as in Section \ref{MSM}, 
with one important difference. When $\tan\beta$ is large, the scalars predominantly couple to 
leptons and down-type quarks. Consequently, the charm and top loops do not contribute in the 
effective couplings to gluons, leading to the reduction in 
the cross-sections for $SS \to gg, \pi\pi$ by a factor of $O(1/N_H^2)\sim 0.1$,
where $N_H$ is the number of heavy quark flavors that convert virtual Higgs to hadrons.
$N_H=3$ in the minimal model, and $N_H=1$ in the 2HDM with large $\tan\beta$.  
The charm threshold also does not offer additional annihilation channels. 
As a result, the values of $\lambda_1$ that would fit the observed abundance 
of dark matter are higher than the corresponding 
values for $\lambda$ in the minimal scalar model studied previously.
The resulting $\kappa$, Figure \ref{2HDM:BR1}a, clearly corresponds to 
a strong interaction regime below WIMP masses of 500 MeV.

\subsection{$\lambda_2$ dominant}

\par
The case in which the scalars couple predominantly to $H_u$ produces results 
similar to the minimal model of Section \ref{MSM}. In the large $\tan \beta$ limit 
$v_u \approx v_{SM}$ and therefore the fine-tuning of the scalar 
mass is significant, while the branching ratio for 
$B \to K + SS$ and constraints on $\lambda_2$ are not 
expected to be significantly different from the minimal model with one Higgs doublet. 

\subsection{$\lambda_3$ dominant}

\begin{figure}
\psfig{file=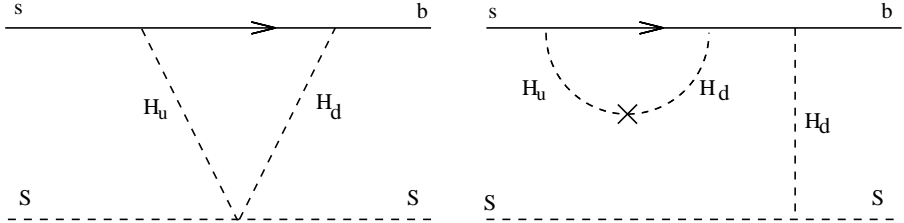,width=12cm,angle=0}
\caption{\label{figure:l3diagrams}Diagrams contributing to the decay $b \to s +SS$ in   the 2HDM plus 
scalar dark matter model when $\lambda_3$ is dominant and $\tan\beta$ is large. }
\end{figure}

\par
The third possibility is that the dark matter scalars couple to the $H_uH_d$ 
combination of the two Higgs fields. In this case the mass of the scalar is

\begin{displaymath}
m_S^2 = 
m_0^2 - \frac{ \lambda_3 v_{EW}^2 }{\tan \beta}
\end{displaymath}

\noindent
which in the large $\tan \beta$ limit requires less fine-tuning to produce sub-GeV 
WIMPs than in the minimal scalar model.

\par

We find that in this model the branching ratio of WIMP pair production in 
$B$-meson decays is suppressed. 
It turns out that two $b\to s +SS$ diagrams  that have $\tan \beta$ enhancement
shown in Figure \ref{figure:l3diagrams} exactly cancel each other 
so that $C_{DM} \sim O(\tan^0\beta)$.


\par
At the same time, the abundance constraint is similiar to the $\lambda_1$ dominant case, 
except that the scalar annihilation cross-section is enhanced by a 
factor of $\tan^2 \beta$, resulting in  $\lambda_3$ being $\tan\beta$ times smaller than 
$\lambda_1$ from the previous example. A combination of the enhanced annihilation cross section and cancellation of 
$b \to s + SS$ amplitude at leading order in $\tan \beta$,  results in  $\tan^{-2}\beta\sim O(10^{-3}-10^{-4})$ 
suppression of $b \to s + SS$ branching ratio relative to previous examples. Consequently, 

\begin{equation}
{\rm Br}_{B \to K + SS} \ll {\rm Br}_{B \to K + \bar \nu \nu} ~~~~~{\rm for ~ \lambda_3~ dominant}
\end{equation}

\noindent
and therefore this model cannot be constrained by $B$-decays. 
It is then clear that a generic case with both couplings, $\lambda_1$ and $\lambda_3$,
being important can create a large range for ${\rm Br}_{B \to K + SS}$
even for a fixed value of $m_S$, and therefore 
the model of scalar dark matter coupled to 2HDM is far less predictive 
than the minimal model.

It also interesting to note that the elastic scattering of dark matter on 
nucleons is enhanced by $\tan^2\beta$ when $\lambda_3$ is dominant, and therefore 
this model will indeed create a measurable signal for detectors 
with low recoil energy threshold such as CRESST. Detailed analysis 
of such constraints goes outside the scope of the present paper.

\section{\label{Fermion} Fermionic Dark Matter}

\par
In the previous two sections the WIMPs were presumed to be scalar fields. 
This is common in models of light dark matter, as the Lee-Weinberg limit on the 
WIMP mass is in general lower for scalar WIMPs. However fermionic dark matter 
could still be light if its annihilation cross section is enhanced.
 In this section we will present some simple models of fermionic 
dark matter, and review the constraints from $B$-decays.

\par
The simplest renormalizable model of fermionic dark matter is the 
analogue of the scalar model given in Eq \ref{tsdm},

\begin{equation}
\begin{split}
-{\cal L}_f &= \frac{m_{\chi}}{2} \overline{\chi} \chi+\frac{m_U^2}{2} U^2+ 
\mu U \overline{\chi} \chi   + \eta U^2 H^{\dag} H
\\&\\ &= \frac{m_{\chi}}{2} \overline{\chi} \chi+\frac{m_u^2}{2} U^2+ 
\mu U \overline{\chi} \chi   + \eta v_{ew} w U h + \lambda_U U^4
\end{split}
\end{equation}

\noindent
where $\chi$ is a Majorana fermion and $H$ is the Standard Model Higgs field. 
As before, $w \equiv <U> \sim O (v_{ew})$. However the annihilation 
cross-section, at freeze out, in this model is suppressed relative 
to the scalar model by \footnote{There is a possibility of producing light 
fermionic WIMPs in this model if $m_u  \ll v_{ew}$, and $\eta \ll \mu < 1$, 
though such a model would require significant fine-tuning to avoid being detected 
in direct searches of light(er) Higgs. }

\begin{displaymath}
\frac{\sigma_{fermion}}{\sigma_{scalar}} \sim \frac{m_{\chi}^2 v_{rel}^2}{m_u^2} \sim O(10^{-5})
\end{displaymath}

\begin{figure}
\psfig{file=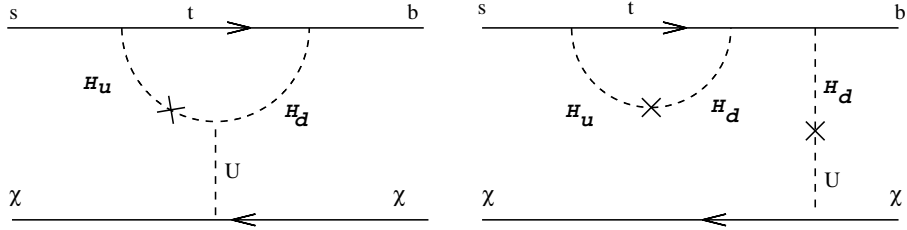,width=12cm,angle=0}
\caption{\label{diagrams:Fermion2} The Feynman diagrams which contribute to 
$b \to s + E\!\!\!\!/$ in the 2HDM plus fermionic WIMP model. }
\end{figure}

\noindent
for  $m_{\chi} \sim O(1\;GeV)$ and $m_u \sim O(v_{ew})$. As a result of this 
suppression, the fermions cannot annihilate efficiently in the early 
Universe and therefore the couplings would have to be non-perturbative to explain dark matter.

\par
The second model we will consider is the analogue of the model presented in 
Section \ref{2HDM}. For the model to be renormalizable, the Majorana fermions 
cannot couple directly to the higgs fields, but must instead couple through an intermediate scalar.

\begin{equation}
\begin{split}
- {\cal L} =&\frac{m_0^2}{2} \overline{\chi}\chi+\frac{m_U^2}{2} U^2 + 
\mu U \overline{\chi}\chi + \lambda_1 U^2 (|H_d^0|^2 + |H_d^-|^2) \\& + 
\lambda_2 U^2 (|H_u^0|^2 + |H_u^+|^2)  + \lambda_3 U^2 (H_d^-H_u^+ - H_d^0 H_u^0)+ \lambda_U U^4
\end{split}
\end{equation}

\begin{figure}
\begin{center}
\psfig{file=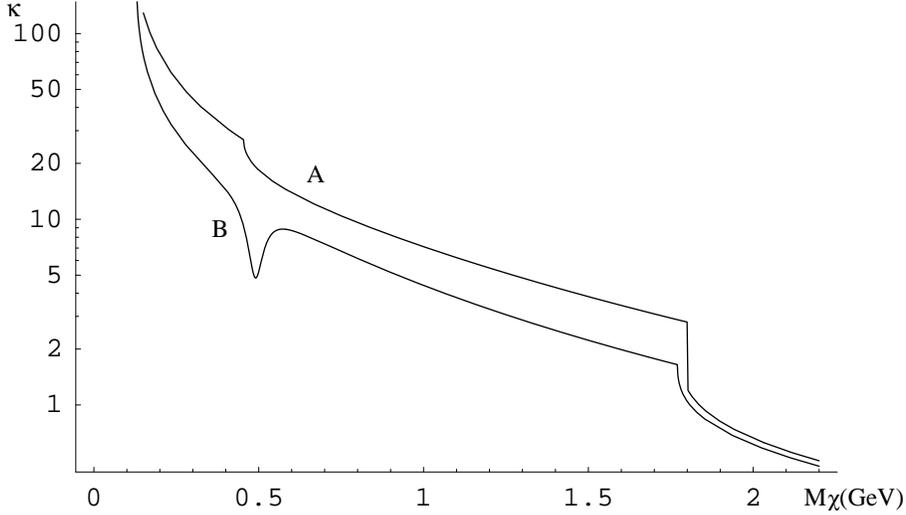,width=12cm,angle=0}
\caption{\label{figure:2HDMchi} Abundance constraints on $\kappa$ in the 2HDM plus fermionic WIMP model. 
It is expected that most of the parameter space near curve B is also excluded, 
however a detailed calculation of the annihilation cross-section 
$\sigma_{\tilde{\chi} \tilde{\chi} \to 3 \pi}$ is beyond the scope of this paper, 
and as such we have utilized the perturbative formula for production of gluons and unbound quarks. 
However except for possible resonances the decay is expected to be significantly smaller, 
and therefore the abundance constraints would require the coupling constant to be larger. }
\end{center}
\end{figure}

\noindent
As in the previous model, the inclusion of fermions in this model 
requires an increase in the coupling constants by a factor of $\sim m_u^2/(w m_{\chi})$, 
with $w = <U>$, to produce the same abundance of dark matter. 
However in the special case of $\lambda_3 \gg \lambda_1, \lambda_2$ the annihilation cross 
section is also enhanced by a factor of $\tan^2 \beta$, and so $\chi$ could be light 
without requiring large $\lambda_3$. The abundance constraints are given in 
Figure \ref{figure:2HDMchi}, with 

\begin{displaymath}
\kappa^2 \equiv 4 \lambda_3^2 \mu^2 \left( \frac{v_{sm} w}{m_u^2} \right)^2 
\left( \frac{100 \; GeV}{M_H} \right)^4 \left(\frac{\tan \beta}{100} \right)^2.
\end{displaymath}

\noindent
As in the $\lambda_3$ dominant case presented in Section \ref{2HDM}, 
the $O(\tan \beta)$ contributions (see Figure \ref{diagrams:Fermion2}) 
to the process $b \to s+\overline{\chi}\chi$ cancel and as a result 
\begin{displaymath}
{\rm Br}(B \to K + \overline{\chi} \chi)) \ll {\rm Br}(B \to K + \overline{\nu} \nu).
\end{displaymath}

\par
The final model we will consider is a Majorana fermion $\chi$ coupled to higgs-higgsino pairs.
By "higgsinos" we mean an $SU(2)\times U(1)$ charged fermionic fields with the same quantum 
numbers as the Higgs fields but do not impose supersymmetry requirements on the size of 
the couplings. By the same token $\chi$ can be called the neutralino. 
The relevant part of the Lagrangian is given by

\begin{displaymath}
- {\cal L}_f = \frac{1}{2} M \overline{\psi} \psi + \mu 
\overline{\tilde{H}_d} \tilde{H}_u + \lambda_d \overline{\psi} \tilde{H}_d H_d + 
\lambda_u \overline{\psi} \tilde{H}_u H_u,
\end{displaymath}

\noindent
where $M \ll \mu, \lambda_u v_u$, and $\tan \beta $ is large. In this model the dark 
matter candidate is taken to be the lightest mass eigenstate,

\begin{equation}
\begin{split}
\chi &= - \psi \cos \theta + \tilde{H}_d \sin \theta  \quad \quad \sin^2 
\theta \equiv \frac{\lambda_u^2 v_u^2}{\lambda_u^2 v_u^2 +\mu^2}
\\&\\
m_1 &= M \left( 1 - \frac{\lambda_u^2 v_u^2}{\lambda_u^2 v_u^2 + \mu^2} \right)
\end{split}
\end{equation}

\noindent
It should be noted that this model is constrained by the $Z$-boson invisible decay width, 
and requires $\sin^2 \theta \lesssim 0.15$. 

\begin{figure}
\begin{center}
\psfig{file=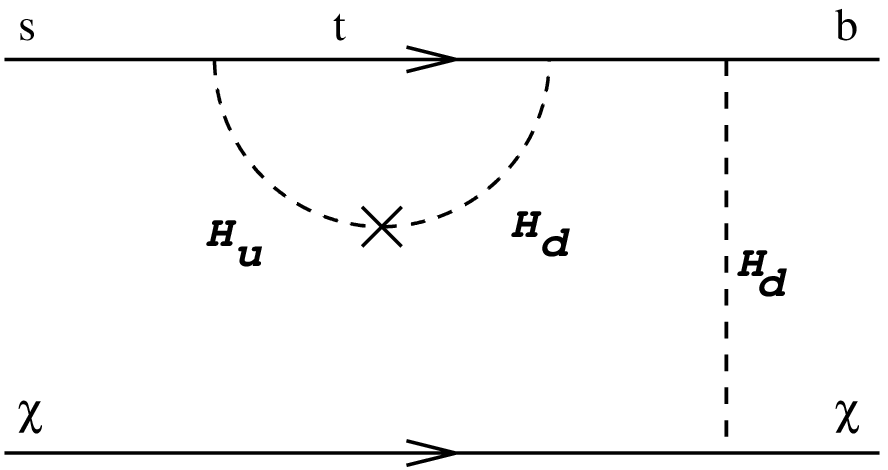,width=7cm,angle=0}
\end{center}
\caption{\label{diagram:Neutralino} Feynman diagram for the 
process $b \to s + \overline{\chi} \chi$ in the final model.}
\end{figure}

\begin{figure}
\begin{center}
$\begin{array}{c}
\psfig{file=kappa-2HDM3.eps,width=12cm,angle=0} \\ (a)\\
\psfig{file=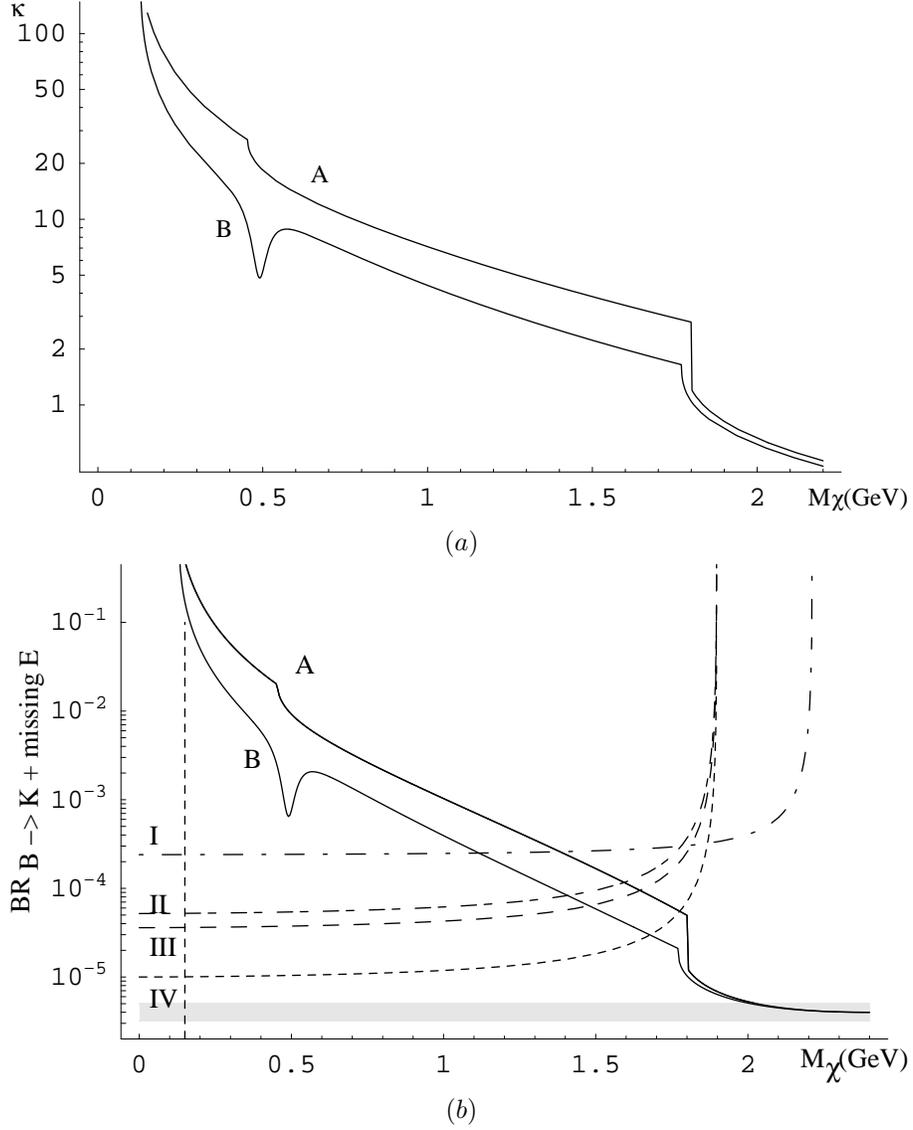,width=12cm,angle=0} \\ (b)
\end{array}$
\end{center}
\caption{\label{figure:fermion3}(a) Constraints on $\kappa$ from the 
observed dark matter abundance. (b)Branching Ratios for $B \to K+missing \; energy$ 
in the neutralino model.  Current limits from \babar,\ BELLE, and 
CLEO are indicated as in Figure \ref{figure:BranchRatio}b. 
The Standard Model predictions are given in the gray box. }
\end{figure}

\par
The $b \to s$ transition in this model is calculated from a single diagram, 
given in Figure \ref{diagram:Neutralino}. The effective Lagrangian is 


\begin{equation}
{\cal L}_{b\rightarrow s E\!\!\!\!/} = \frac{1}{2}C_{DM} m_b\bar s_L b_R \overline{\chi}\chi -
C_\nu \bar s_L \gamma_\mu b_L \bar \nu \gamma_\mu \nu+(h.c.)
\label{bsE2}
\end{equation}

\noindent
with Wilson coefficient
\begin{equation}
C_{DM} = \frac{V_{ts}^* V_{tb} \tan \beta}{32 \pi^2 v_{sm}^3} 
\left( \frac{\lambda_d \lambda_u v_u \mu}{\lambda_u^2 v_u^2 + \mu^2} \right) \frac{a_t \ln a_t}{1-a_t},
\end{equation}

\noindent
where $a_t \equiv m_t^2/M_H^2$, and the branching ratio is

\begin{equation}
BR_{B \to K + E\!\!\!\!/} = 4.0\times10^{-6}+9.8 \times 10^{-5} \kappa^2 
\left( \frac{\ln a_t}{1-a_t} \right)^2 F(m_1),
\end{equation}

\noindent
where

\begin{displaymath}
\kappa^2 \equiv \left( \frac{\lambda_d \lambda_u v_u \mu}{\lambda_u^2 v_u^2 + \mu^2} \right)^2 
\left( \frac{100 \; {\rm GeV}}{M_H} \right)^4 \left( \frac{\tan \beta}{100} \right)^2.
\end{displaymath}

\noindent
The phase space integral for fermions is defined as

\begin{equation}
F(m_{\chi})=\int_{\hat s_{min}}^{\hat s_{max}}\!\!\!\!\!\!\! 
f_0(\hat{s})^2 (\hat{s}-2 m_{\chi}^2/M_B^2)I(\hat s,m_{\chi})~d\hat s ~\left[
\int_{\hat s_{min}}^{\hat s_{max}}
f_0(\hat{s})^2 \hat{s} I(\hat s,0)~d\hat s~\right ]^{-1}.
\label{fermiphase}
\end{equation}

\noindent
The cosmological abundance and the branching ratio are plotted in 
Figure \ref{figure:fermion3} with the current experimental limits. 
Although it appears that the effects of dark matter in this 
model could be observed, the abundance constraints require $\kappa$ to be in the 
non-perturbative region for $m_{\chi} \lesssim 1  $ GeV. Furthermore, constraints on 
$\lambda_u$ from $Z$-decays and on $M_{H_d}$ require the original coupling constant 
$\lambda_d $ to be a few orders of magntude larger than $\kappa$ and therefore 
the model is non-perturbative for $m_{\chi} \lesssim 2$ GeV. Heavier WIMPs may 
still be produced in $B$-decays, though the branching ratio is significantly 
smaller than the Standard Model signal.

\section{NMSSM with light dark matter}

\par
One of the most popular dark matter candidates is the lightest 
neutralino present in supersymmetric models. The existence of supersymmetry 
is well motivated in particle physics, and the cosmological abundance of 
neutralinos is comparable to the dark matter abundance for a large region 
of parameter space. In this section we will demonstrate that light neutralinos, 
if kinematically allowed, can be produced in $B$-meson decays and will provide a 
significant contribution to the branching ratio Br$(B \to K + 
E\!\!\!\!/)$.

\par
In discussions of supersymmetric dark matter, it is common to work with the minimal 
supersymmetric standard model (MSSM). However in that case the lightest neutralino 
is expected to be heavier than a few tens of GeV and 
therefore too heavy to produce in $B$-meson decays. 
Although it is possible to lower this limit to 6 GeV \cite{Bottino:2003iu}, it requires 
extreme tuning of the parameters, while sub-GeV WIMPs seem to be impossible even with the 
fine tuning. However it is possible to produce sub-GeV neutralinos in extensions of MSSM
by introducing new resonances significantly lighter than $M_W$.

\par
In this section we present a case study of NMSSM with light neutralino dark matter. 
The lightness of $\chi$ in this model is made possible, without significant fine-tuning \cite{Dermisek:2005ar}, by one of the pseudoscalar Higgs bosons $a$
being chosen to have a mass of a few GeV. A comprehensive analysis of dark matter in this model was attempted 
in Ref. \refcite{NMSSM}, while the constraints from $B$-physics were considered in Ref. \refcite{Hiller:2004ii}. 
None of these papers, however, considers the decays of $B$-mesons with missing energy.
As mentioned in the previous section, the neutralino annihilation cross section in the non-perturbative 
QCD regime has not been adequately addressed. A proper account of $m_\chi <1$ GeV would require 
the calculation of $a\to gg \to PPP$ and $ a\to qq \to PPP$, where $P$ is a pseudoscalar meson such as 
$\pi$ and $K$, plus an extensive scan over the NMSSM
parameter space, which we will not attempt here. Instead, we choose to analyze 
the domain of neutralino masses of $m_\mu < m_\chi < 1.5 m_\pi$ 
where the annihilation proceeds primarily via muons in the final state, and is free from hadronic uncertainties:
\begin{equation}
\sigma_{ann}v_{rel} \simeq \frac{1}{2\pi}\left(\frac{\lambda_{a\chi}\sin\gamma\tan\beta}{m_a^2}\right)^2
\left(\frac{m_f}{v_{EW}}\right)^2m_\chi^2 \sqrt{1-m_\mu^2/m_\chi^2}.
\label{sigma_ann}
\end{equation}
In this formula, the coupling constants $\lambda_{a\chi}$ and mixing angles 
$\gamma$ originate from the tree-level 
interaction Lagrangian of $a$ with the SM fermions and neutralinos,
\begin{eqnarray}
-{\cal L}_{int} = \frac12\lambda_{a\chi } a\bar\chi i \gamma_5 \chi 
+ \sin\gamma\tan\beta \sum \frac{ m_f}{v_{EW}} a\bar f i\gamma_5 f.
\end{eqnarray}
Here the summation goes over the charged leptons and down-type quarks,  $\gamma$ 
parametrizes the admixture of Im$H_d$ in $a$, and $\tan\beta$ is assumed to be a large parameter. 
Specific relations between $\gamma$, the neutralino-pseudscalar
coupling $\lambda_{a\chi}$ and the parameters of the fundamental NMSSM Lagrangian 
can be worked out, but are of no interest in this discussion since they bear no consequences
for our predictions. 

\begin{figure}
\begin{center}
\psfig{file=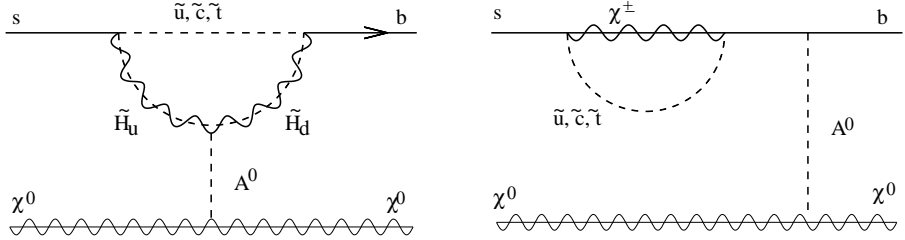,width=12cm,angle=0}
\end{center}
\caption{\label{diagrams:NMSSM}Additional Feynman diagram for the process 
$b \to s + \overline{\chi} \chi$ in the NMSSM model.}
\end{figure}

The SUSY diagrams that can play an important role in the $b\to s + \chi\chi$ decay are shown in 
Figure \ref{diagrams:NMSSM}.
 These are two Higgsino-stop exchange diagrams that have 
different parametric $\tan\beta$ and $\gamma$ dependence. 
They give the folowing Wilson coefficient in the effective Lagrangian defined in (\ref{bsE2}),
\begin{equation}
C_{DM} = \frac{V_{ts}^* V_{tb} g_W^2 x_t\tan \beta}{32 \pi^2 v_{EW} } 
\left(\frac{\lambda_{a\chi}\sin\gamma\tan\beta}
{m_a^2}\right) I_{SUSY}.
\label{susycdm}
\end{equation}
In this expression, we  adhere to the
minimal flavour-violation scheme of the soft-breaking sector
and treat the stop mixing as a mass insertion. The loop function 
$I_{SUSY}$ depends on the trilinear soft-breaking parameter $A_t$, masses of left- and right-handed stops,
which for simplicity can be chosen equal, $m_{LL}^2\simeq m_{RR}^2\simeq m_{\tilde t}^2$,
and $\mu$ parameter: 
\begin{eqnarray}
\label{Isusy}
I_{SUSY} = \frac{A_t \mu}{m_{\tilde t}^2}
\left(\frac{1-a_{\tilde t}+a_{\tilde t}\ln a_{\tilde t}}{(1-a_{\tilde t})^2}\right)
\left[1 + \frac{\lambda_{a\tilde H}}{\tan\beta\sin\gamma}\right],
\end{eqnarray}
where $a_{\tilde t} = \mu^2/m_{\tilde t}^2$. Note that in the context of NMSSM the Higgsino-pseudoscalar coupling
$\lambda_{a\tilde H}$ and $\mu$-parameter are  often chosen to to be 
related via the singlet's vev. 
It is important that $I_{SUSY}$ does not decouple even in the limit of large SUSY masses, provided that the 
relative size of $A_t$, $\mu$ and $m_{\tilde t}$ remains fixed. When all SUSY masses are equal, 
$A_t = \mu = m_{\tilde t}$, the expression in front of the square bracket in (\ref{Isusy}) becomes 1/2,
which we would consider as a natural value for $I_{SUSY}$.  Notice an additional power of $\tan\beta$ in 
the expression for $C_{DM}$ (\ref{susycdm}), a very well-known enhancement in the SUSY models \cite{tanbeta-enh,tanbeta-enh2,tanbeta-enh3}. 
By introducing an effective coupling $\kappa$,
\begin{equation}
\kappa ^2 \equiv
\left(\frac{\tan\beta}{10}\right)^2\left(\frac{10~{\rm GeV}}{m_a}\right)^4 (\lambda_{a\chi}\sin\gamma)^2,
\end{equation}
we express the $B$-decay misisng energy signal as
\begin{equation}
{\rm Br}_{B \to K + E\!\!\!\!/} = 4.0\times10^{-6}+
0.98 \kappa^2 \left( \frac{\tan\beta}{10} I_{SUSY}\right)^2 F(m_\chi)
\end{equation}
with
\begin{equation}
F(m_{\chi})=\int_{\hat s_{min}}^{\hat s_{max}}\!\!\!\!\!\!\! 
f_0(\hat{s})^2 
\hat{s} 
I(\hat s,m_{\chi})~d\hat s ~\left[
\int_{\hat s_{min}}^{\hat s_{max}}
f_0(\hat{s})^2 \hat{s} I(\hat s,0)~d\hat s~\right ]^{-1}.
\end{equation}


Using the freeze-out abundance to constrain $\kappa$, 

\begin{equation}
\kappa \lesssim \frac{2.4 ~{\rm GeV}}{m_{\chi}(1-m_{\mu}^2/m_{\chi}^2)^{1/4}}
\end{equation}

\noindent
and imposing the experimental bound of 
Br$(B^+ \to K^+ + {E}\!\!\!\!/) < 3.5\times 10^{-5}$ 
on the branching ration, 
we arrive at the following constraint on $I_{SUSY}$,
\begin{equation}
\left( \frac{\tan\beta}{10} I_{SUSY}\right) < 3.0 \times 10^{-4} ~~~~{\rm for}~ m_\chi = 150~{\rm MeV}
\end{equation}
This is a requirement on $I_{SUSY}$ to be suppressed relative to its natural value by 
three orders of magnitude. 
We conclude that unless an unnatural cancellation of $I_{SUSY}$ occurs, the possibility of a 
$\sim 100$ MeV neutralinos in NMSSM is excluded. 
However, even if the SUSY contribution vanishes for some reasons, 
the next-to-leading order in $\tan\beta$ contribution mediated by 
charged Higgs loops is still capable of producing an observable signal, 
which from existing bounds still excludes a $\sim 100$ MeV neutralino.

\section{\label{experiment} Experiment}
The decays of the type $B^+ \to K^+ + {E}\!\!\!\!/$ discussed here
involve missing energy but provide no constraint on the associated
missing mass.  This is problematic experimentally, even for a
hypothetical detector with full solid angle coverage, unless one of the
two $B$ mesons in the event is fully reconstructed.  Similarly, it
is hard to imagine that a sensitive search can be done without the
well-constrainted initial state provided by experiments at the
threshold for $B\overline{B}$ production.  This discussion will
therefore focus on experiments at the $\Upsilon(4S)$.

The $B$
reconstruction can be done using either hadronic or semileptonic
decays.  After the decay products of one $B$ are removed from
consideration, signal events should contain only a single kaon while
background events will generally contain additional charged or neutral
particles.  In reality, beam-related backgrounds tend to leave additional
energy in the electromagnetic calorimeter and interactions of particles
from the fully reconstructed $B$ with material in the detector can
lead to additional, spurious charged or neutral particles.  The
selection criteria balance the loss of efficiency from strict vetoes
on additional activity against the suppression of background.  After
removing particles associated with the fully reconstructed $B$,
no charged tracks are allowed beyond the identified kaon, and
the neutral electromagnetic energy deposition is
restricted to be less than 200-300~MeV.  
The background is negligible
for $K$ energies near $m_B/2$ and rises steadily toward lower $K$
momenta.  As a result, the experiments place a requirement on the minimum
accepted $K$ momentum; this is further discussed below.

The CLEO measurement reconstructs $B$ mesons in hadronic decays
and requires $p_K>0.7$~GeV, providing sensitivity to dark matter
particle masses up to 2.1~GeV and placing a limit of 
Br$(B^+ \to K^+ \nu\overline{\nu})<24\times 10^{-5}$ for the SM decay.  
The \babar\ and BELLE measurements each require $p_K>1.2$~GeV, thereby
restricting sensitivity to dark matter particle masses below 1.9~GeV.
\babar\ reconstructs $B$ mesons in both hadronic and semileptonic
modes with comparable sensitivity in each, and determines
Br$(B^+ \to K^+ \nu\overline{\nu})<5.2\times 10^{-5}$ based on
a sample of $88\times 10^6$ $\Upsilon(4S)\to B\overline{B}$ decays.
BELLE uses only events with one $B$ reconstructed in a hadronic
decay mode and find the preliminary result
Br$(B^+ \to K^+ \nu\overline{\nu})<3.5\times 10^{-5}$ based on
a sample of $275\times 10^6$ $\Upsilon(4S)\to B\overline{B}$ decays.

The expected background from normal $B$ decays with
two or more missing particles is several times higher than the
signal expected from the SM process with neutrino pair.  This background has contributions
from semileptonic decays (e.g. $\bar{B}\to D\ell\overline{\nu}$ with 
$D\to K^+\ell^-\nu$ or $D\to K^+X$) where the the detectable particles
other than the $K^+$ lie outside the detector
acceptance, as well as from hadronic $B$ decays involving one or more
$K^0_L$, which have a non-negligible probability of traversing the
electromagnetic calorimeter without depositing significant energy.  
The latter decays in some cases still have poorly known branching 
ratios, but these can be measured (or more stringently bounded) with
larger $B$-factory data sets.  The sensitivity can 
be expected to improve as $1/\sqrt{N}$ as more data are analyzed.

The decay $B^0\to K^0 + {E}\!\!\!\!/$ doesn't
add much sensitivity due to the smaller reconstruction efficiency
for both the $B^0$ (about 1/2 that of $B^+$) and the $K^0$ (only
1/3 decay to $\pi^+\pi^-$).  
The decays $B\to K^* + {E}\!\!\!\!/$ have sensitivity comparable to
$B^+ \to K^+ + {E}\!\!\!\!/$ (the branching ratios for both the
SM and New Physics transitions are higher due to the higher probability
of producing a $K^*$ in the fragmentation of the $s$ quark, offsetting
the lower experimental efficiency).  
The $B \to K^* + {E}\!\!\!\!/$ channel should be sensitive to
dark matter masses only slightly smaller than those explored in
the $B^+ \to K^+ + {E}\!\!\!\!/$ channel, allowing for
an improvement in overall sensitivity to light dark matter.
This, coupled with the $\sim 1$~ab${}^{-1}$ data samples that will
soon be available from the $B$ factories, should push the sensitivity
to missing energy processes to $\sim 2$ times the level of the SM branching ratio.

\section{Conclusions}

\par
In conclusion, we have demonstrated that $B$-decays with missing 
energy provide a viable method for exploring light dark matter models. 
This alternative to dedicated underground experiments has considerable sensitivity to 
light WIMPs, and can probe dark matter using existing experimental results. 
While this result has been previously demonstrated in Ref \refcite{Bird:2004ts}, 
this paper extends the result to include several different models. 

\par
In particular it has been demonstrated that the 2HDM+scalar model, in the case of 
$\lambda_1$ dominant, can satisfy the abundance constraint with Br$(B^+ \to K^+ + SS) > {\rm Br}
(B^+ \to K^+ + \bar{\nu} \nu) $.  This model has an additional benefit, as it
requires only minimal fine tuning to achieve sub-GeV WIMPs in the large tan $\beta$
 limit. Data from \babar\ and BELLE excludes the parameter space below 
$m_S \sim 1.3 \; $GeV, while future measurements are expected to 
probe up to $m_S \sim 1.9 \; $GeV. The case of $\lambda_3$ dominant also provides sub-GeV WIMPs 
with minimal fine-tuning. However in that case the branching ratio  Br$(B^+ \to K^+ + SS)$ 
is suppressed by factor of $\tan^2 \beta$ relative to the $\lambda_1$ dominant case, 
and therefore is significantly smaller than the Standard Model signal. In this case, only the 
direct searches with the low recoil energy threshold has a chance of detecting these WIMPs. 

\par
Furthermore we have demonstrated that it is possible to search for fermionic dark
 matter in $B$-decays, though in the models presented the couplings are 
required to be non-perturbative to meet the abundance constraint for 
$m_{\chi} \lesssim 2  $ GeV. While it is possible that heavier 
fermions could be produced in $B$-decays, the
 branching ratio for such a process is significantly smaller than the 
Standard Model predictions. It is also possible to lower the 
Lee-Weinberg limit in these models
  by allowing other light particles which are very weakly coupled to 
the Standard Model, such as in the NMSSM with one light pseudoscalar Higgs. 
In such a model, it would be possible to have weak couplings and 
still produce an observable increase in 
Br$(B \to K + missing \; energy)$. Our analysis shows that indeed $O(100)$ MeV 
neutralinos of NMSSM are generically excluded by the $B$ decay data, while heavier 
(above 1 GeV) neutralinos would probably be compatible with existing constraints. 

\par
Except for the case of NMSSM, we have limited the discussion to models which do not include new
 forces below the electroweak scale. Models with new resonances below $M_W$ 
can be explored using $B$-decays, but in addition
they might be well constrained by existing experiments and therefore should be introduced with caution. 
Such models can be analyzed in exactly the same manner as presented in this paper.



\end{document}